\documentclass[useAMS,usenatbib]{mn2e}
\usepackage{graphicx,psfig}
\usepackage{natbib}
\usepackage{txfonts}
\newcommand\apj{ApJ}
\newcommand\apjs{ApJS} 
 
\newcommand\aap{A\&A}

\newcommand\alp{$\alpha$ Centauri}

\def\be{\begin{equation}}
\def\ee{\end{equation}}

\title{Planet formation in the habitable zone of alpha Centauri B}

\author[P. Th\'ebault, F. Marzari, H. Scholl]
{P. Th\'ebault$^{1}$\thanks{E-mail:philippe.thebault@obspm.fr},
F. Marzari$^{2}$, H.Scholl$^{3}$\\
$^{1}$Stockholm Observatory, Albanova Universitetcentrum,
SE-10691 Stockholm, Sweden, and\\ 
Observatoire de Paris, Section de Meudon, F-92195 Meudon Principal Cedex,
France\\
$^{2}$Department of Physics, University of Padova, Via Marzolo 8,
35131 Padova, Italy\\
$^{3}$Laboratoire Cassiop\'ee, Universit\'e de Nice Sophia Antipolis, CNRS, Observatoire de la C\^ote d'Azur, B.P. 4229,
F-06304 Nice, France
}

\begin{document}

\date{Draft version \today}

\maketitle

\begin{abstract}
Recent studies have shown that alpha Centauri B might be, from 
an observational point of view,
an ideal candidate for the detection of 
an Earth-like planet in or near its habitable zone 
(0.5-0.9AU).
We study here if such habitable planets can form, by numerically
investigating the planet-formation stage which is probably the most sensitive
to binarity effects: the mutual accretion of km-sized planetesimals.
Using a state-of-the-art algorithm for computing the impact velocities
within a test planetesimal population, we find that planetesimal
growth is only possible, although marginally,
in the innermost part of the HZ around
0.5AU. Beyond this point, the combination of secular
perturbations by the binary companion and gas drag drive the 
mutual velocities beyond the erosion limit.
Impact velocities might later decrease during the gas removal phase,
but this probably happens too late for preventing most km-sized
objects to be removed by inward drift, thus preventing accretion
from starting anew. 
A more promising hypothesis is that the binary formed in a crowded 
cluster, where it might have been wider in its initial stages, when 
planetary formation was ongoing. We explore this scenario and
find that a starting separation roughly 15\,AU wider,
or an eccentricity 2.5 times lower than the present ones
are required to have an accretion-friendly environment in the whole HZ.

\end{abstract}


\section{Introduction} \label{sec:intro}

The study of planet formation in binary stars
is an important issue, as approximately $\sim 20\%$ of all
detected exoplanets inhabit such systems \citep{desi07}.
Even if most of these binaries are very wide, at least 3 planets
have been found in binaries of separation $\sim$20 AU, the most
interesting of them being the $M \geq 1.6M_{Jup}$ planet at 2.1AU
from the primary of the $\gamma$ Cephei system, which has been
the subject of several planet-formation studies
\citep{theb04,kley08,jang08}.

However, the system which has been most thoroughly investigated
is one where no planet has been detected so far: alpha Centauri
\footnote{\citet{endl01} rule out the presence of a
$M\geq2.5\,M_{Jupiter}$ planet at any radial distance}.
\citet{holw97} have shown that, for the coplanar case, the $<3\,$AU
region around $\alpha\,$Cen\,A can harbour planets on stable orbits, while 
\citet{barb02} and \citet{quin02,quin07} have found that the 
final stages of planetary formation, the ones leading from large
planetary embryos to the final planets, are possible in
the inner $<2.5\,$AU region.
However, \citet[][hereafter TMS08]{theb08} showed that the
earlier stage leading, through runaway growth in a dynamically quiet
environment \citep[e.g.][]{liss93}, from kilometre-sized planetesimals to the
embryos is much more affected by the companion star's perturbations.
Indeed, the coupled effect of these secular perturbations and drag due
to the gas in the nebulae leads to a strong orbital
phasing according to planetesimal sizes. This differential phasing
induces high collision velocities for any impacts between objects
of sizes $s_1 \neq s_2$, which can strongly slow down or even halt
the accretion process. As a result,
the region beyond 0.5\,AU (0.75\,AU for the most extreme cases explored)
from the primary is hostile to mutual accretion of planetesimals, and thus
to the in situ formation of planets. The {\it presence} of planets
in these regions, however, cannot be ruled out, in the light
of possible additional mechanisms not taken into account
in the simulations, such as a change of the dynamical environment
once gas has been dispersed, or outward migration of planets formed
closer to the star, or a change in the binary's separation
during it's early history.

We reinvestigate here these issues for the case of the secondary star
$\alpha$ Centauri B. Indeed, a recent study by \citet{guedes08}
has shown that, in addition to the final embryos-to-planets stage
being possible within $\sim 2.5\,$AU from it, this star would
be an almost perfect candidate for the detection of a potential
terrestrial planet by the radial velocity method. This is mainly
because of its great proximity, but also because it is
exceptionally quiet. Moreover, because of its lower luminosity,
its habitable zone (HZ) lies much closer, 0.5 to 0.9\,AU
\citep{guedes08}, than 
that of $\alpha\,$Cen\,A \citep[$\sim$1 to 1.3 AU, e.g.][]{barb02}. 

In this letter, we first numerically study the dynamics of the
planetesimal accretion phase in a similar way as for
$\alpha\,$Cen\,A, by esimating the extent of the
accretion-friendly region around the central star (section II).
We then push our studies a step further than in TMS08, by
quantitatively exploring two mechanisms which could potentially
increase the odds for planet formation in the HZ (Section III).
Conclusions and perspectives are given in the last Section.

\section{Simulations}

\subsection{model and setup}

We consider a disc of $10^{4}$ test planetesimals extending
from 0.3 to 1.6\,AU from the central star,
starting on quasi-circular orbits such as initial encounter velocities
$\langle \Delta v \rangle <1$ m.s$^{-1}$,
and with a physical radius $s$ in the 1km$<s<$10km range. The binary's
parameters are taken from \citet{pourb02}: semi-major axis
$a_b=23.4\,$AU, eccentricity $e_b=0.52$, $M_A = 1.1 M_{\odot}$
and $M_B = 0.93 M_{\odot}$.
We follow the dynamical evolution of the planetesimal population for
$t=10^4$years, the typical timescale for runaway growth,
under the coupled influence of the companion star's
(here $\alpha\,$Cen\,A) perturbations and gas drag, using the
deterministic code developed by \citet{thebra98}. 
All test planetesimal collisions are tracked with the code's build-in
close-encounter search routine, looking at each time step for all
intersecting particle trajectories with the help
of the classical "inflated radius"
procedure (see section 2.1 of TMS08 for more details).
The gas disc is assumed
axisymmetric with a volumic density profile following
$\rho_{\rm g} = \rho_{\rm g0}\,r_{(\rm{AU})}^{q}$, where we consider as a nominal
case the Minimum Mass Solar Nebula (MMSN), with $q=-2.75$ and
$\rho_{\rm g0}=1.4\times10^{-9}$g.cm$^{-3}$ \citep{haya81}, exploring
$\rho_{\rm g0}$ and $q$ as free parameters in separate runs.

All mutual encounters and corresponding encounter velocities are tracked
and recorded in order to derive $\langle \Delta v \rangle_{s_1,s_2}$ matrices
for all possible impactor pairs of sizes $s_1$ and $s_2$.
These velocity values are then interpreted in terms of accreting or eroding
encounters by comparing them, for each impacting pair, to 2 threshold
velocities \citep[for more details, see][]{theb06,theb08}:
\begin{itemize}
\item $v_{esc(s_1,s_2)}$, the escape velocity of the impacting pair. This velocity
gives approximately the limit below which single star-like
runaway accretion is possible
(when accretion rates are enhanced by the gravitational focusing factor).
\item $v*_{(s_1,s_2)}$, the velocity above which impacts preferentially lead
to mass loss rather than mass accretion. This value is obtained from estimates
of the critical threshold energy for catastrophic fragmentation $Q*_{(s_1,s_2)}$.
Given the large uncertainties regarding the possible values for $Q*_{(s_1,s_2)}$,
we consider 2 limiting values, $v*_{(s_1,s_2){\rm low}}$ and 
$v*_{(s_1,s_2){\rm high}}$, corresponding to a weak and hard material assumption
respectively.
\end{itemize}
Finally, at any given location in the system, the global balance
between accreting and eroding impacts is estimated assuming a size
distribution for the planetesimal population, weighting the contribution
of each ($s_1,s_2$) impact by its normalized probability according to
the assumed size distribution. As a nominal case, we assume 
a Maxwellian peaking at the median value $s_{med}=5\,$km,
but other distributions (Gaussians, power laws) are also explored.

\subsection{results: nominal case}

\begin{figure}
\includegraphics[width=\columnwidth]{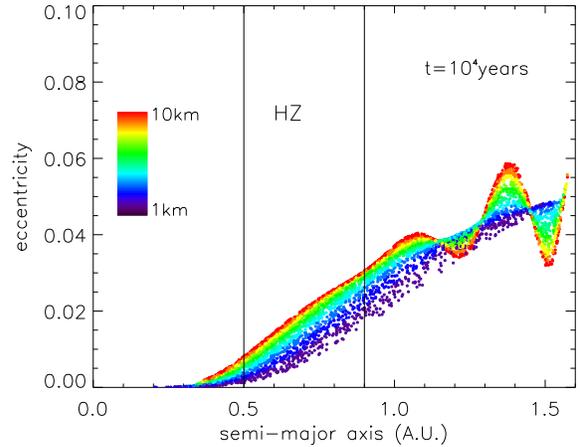}
\caption[]{eccentricity vs. semi-major axis graph at the end
($10^{4}$years) of the nominal case run. The colour scale shows
the different particle sizes. The two vertical lines indicate the
estimated location of the habitable zone.
}
\label{eanom}
\end{figure}

\begin{figure}
\includegraphics[width=\columnwidth]{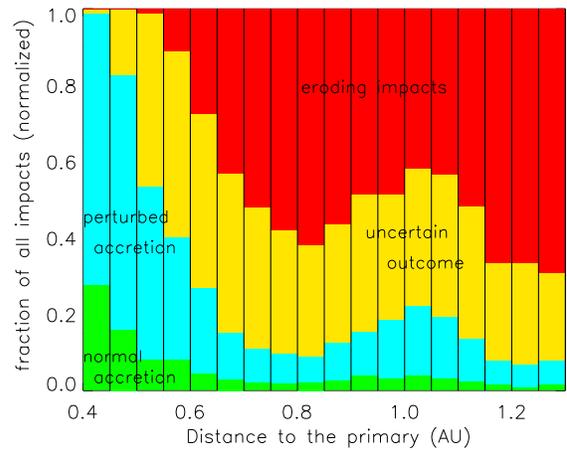}
\caption[]{Relative importance of different types of collision outcomes,
as a function of distance to the primary star ($\alpha\,$Cen\,B), at t=$10^{4}$years
for our nominal case. Each $\Delta v_{(s_1,s_2)}$ for an impact
between bodies of sizes $s_1$ and $s_2$ is interpreted
in terms of eroding impacts (if $\Delta v>v*_{(s_1,s_2){\rm high}}$), 
accreting impacts despite of increased velocities ("perturbed accretion"
for $v_{esc(s_1,s_2)}<\Delta v<v*_{(s_1,s_2){\rm low}}$), or
unperturbed "normal" accretion ($\Delta v<v_{esc(s_1,s_2)}$). For
$v*_{(s_1,s_2){\rm low}}<\Delta v< v*_{(s_1,s_2){\rm high}}$, we do not draw any
conclusions. The contribution of each impact is then
weighted assuming that the size distribution for the planetesimals follows
a Maxwellian centered on 5km (see text for details).
}
\label{histacc}
\end{figure}

The (e,a) graph displayed in Fig.\ref{eanom} clearly illustrates the effect
of differential orbital phasing as a function of sizes. Smaller planetesimals
align to orbits with lower eccentricities than that of the bigger objects.
\footnote{This differential alignment in $e$ is strengthened by the
associated differential alignment of the pericenter angle
\citep[e.g.,][]{theb06}}. Note that in the outer regions (beyond
$\sim 1\,$AU), residual eccentricity oscillations are observed for the
larger bodies. This is because these objects, which are less affected by
gaseous friction, have not had the time yet to fully reach
the equilibrium orbit forced by gas drag.

Fig.\ref{histacc} shows how $\langle \Delta v \rangle_{s_1,s_2}$
resulting from this differential phasing can be
interpreted in terms of accreting versus eroding impacts. This graph
displays the relative weights of the possible types of collision outcomes
as a function of distance to $\alpha\,$Cen\,B, for our nominal case (MMSN gas disc),
assuming that the planetesimal size distribution follows a 
Maxwellian centered on 5km. By comparing
this graph to the equivalent graph for $\alpha\,$Cen\,A (Fig.3 of TMS08), we
see that results are roughly comparable for both stars. Schematically, the region
beyond $\sim$0.5\,AU from the primary is hostile to planetary accretion. 
This similarity of the results is mostly due to the fact that the mass ratio
between the 2 stars is close to 1 ($\sim 0.85$), and that the strongest
perturbations of $\alpha\,$Cen\,A on $\alpha\,$Cen\,B are partly compensated by
the lower Keplerian velocities around that smaller star.
One important difference, however, is that the HZ around $\alpha\,$Cen\,B lies
closer to the star: 0.5 to 0.9 AU \citep{guedes08}. This means that its innermost
part lies at the outer-edge of the accretion-friendly region, thus opening
a possibility, although marginal, for planet formation there.

\subsection{parameter exploration}

\begin{table}
%\begin{center}
\caption[]{Outer limit $r_{acc(out)}$ of the accretion-friendly inner zone,
as defined by the region where the "green" and "blue" areas
make up more than 50\% of the collision outcomes (see Fig.\ref{histacc}),
for test runs exploring different free parameters: 
gas density $\rho_{\rm g0}$ at 1AU, slope $q$ of the 
$\rho_{\rm g} \propto r^{q}$ gas density profile, and
assumed size distribution for the planetesimal population}
\label{explor}
\begin{tabular}{ll}
Set-Up & $r_{acc(out)}$\\
\hline
Nominal case & 0.5\,AU\\
$\rho_{\rm g0}=1.4\times10^{-10}$g.cm$^{-3}$ (0.1xMMSN) & 0.3\,AU\\
$\rho_{\rm g0}=1.4\times10^{-8}$g.cm$^{-3}$ (10xMMSN) & 0.8\,AU\\
$\rho_{\rm g} \propto r^{-2.25}$ ($\alpha$ viscous disc) & 0.45\,AU\\
$\rho_{\rm g} \propto r^{-1.75}$  & 0.4\,AU\\
Size dist.: Gaussian, $\sigma^2=(2\,{\rm km})^{2}$ & 0.55\,AU\\
Size dist.: Gaussian, $\sigma^2=(1\,{\rm km})^{2}$ & 0.6\,AU\\
Size dist.: Gaussian, $\sigma^2=(0.5\,{\rm km})^{2}$ & 1.1\,AU\\
Size dist.: power law $dN \propto s^{-3.5}ds$ & 0.5\,AU\\
\hline
\end{tabular}
%\end{center}
\end{table}

To check the robustness of these results with respect to our choice of
parameters for the nominal case, we firstly explore other possible
gas disc profiles, by varying both the slope $q$ and the
density $\rho_{\rm g0}$ at 1AU. We find
that the only case which leads to a wider accretion-friendly zone is
that of a dense 10xMMSN gas disc, for which this region extends up to
$\sim 0.8\,$AU (Tab.\ref{explor}). 
All other gas-disc cases lead to equally narrow
or even narrower accretion-friendly regions than for our nominal case, except
for a purely academic gas-free case.
Of course, our axisymmetric gas disc assumption is probably a
crude simplification of the real behaviour of a gas disc perturbed by
a companion star. However,
preliminary Hydro+N-body runs performed by \citet{paard08}
show that the build up of an eccentricity for the gas-disc,
altough this mechanism is yet not fully understood,
always lead to $higher$ perturbations and impact velocities in
the planetesimal population.
As a consequence, our axisymmetric disc case might be regarded as a
best-case scenario and our results to be on the conservative side
(for a more detailed discussion and justification for our choice
of an axisymmetric disc, see TMS08).

Another crucial parameter which has been explored is the assumed size
distribution for the planetesimal population, as it is a parameter
poorly constrained from planet-formation models
\citep[see discussion in][]{theb06}. For any "reasonable"
distributions, we do not observe a significant displacement of
the accretion/erosion radial limit. Only a very peaked, almost
Dirac-like Gaussian of variance $\sigma^2=(0.5\,{\rm km})^{2}$ leads to
mostly accreting impacts in the whole HZ (Tab.\ref{explor}).

\section{Discussion}

The results of the previous section show that, as was the case for
$\alpha\,$Cen A, the region allowing km-sized planetesimal accretion
is much more limited than the one allowing the final
stages of planet formation (starting from large embryos):
the accretion-friendly zone only extends up to $\sim 0.5\,$AU
for our nominal case, as compared to the estimated 2.5\,AU for the
embryos-to-planets phase \citep[][ and reference therein]{guedes08}.

The only cases leading
to an accretion-friendly environment in a wider area than for the
nominal run are: a) the gas free case, b) the very-peaked Gaussian size
distribution, and c) the 10xMMSN run. Cases a) and b) are
obviously very unrealistic test runs, but even case c) is probably
here of limited physical significance. Indeed, assuming such a high-mass
disc around a $0.9M_{\odot}$ star appears as a rather extreme hypothesis,
which is not backed by theoretical estimates or observational data
\citep[e.g.][]{andrews07}.
For a binary system, this issue gets even more critical,
as the expected outer truncation of the circumprimary gas disc 
probably depletes it from a substancial amount of it mass reservoir
\citep[e.g.,][]{jang08}, making the 10xMMSN hypothesis even more
questionable, although it cannot be completely ruled out.

Apart from this massive gas disc case, the $r_{acc(out)}\sim0.5\,$AU
limit allows in principle for planetesimal accretion to occur
in the innermost part of the habitable zone, which is estimated
to extend from 0.5 to 0.9AU (as compared to
$\sim1$ to $1.3\,$AU for $\alpha\,$Cen\,A).
However, even if some planetesimal
accretion is possible, it could probably not be what it is
in the standard planet formation scenario. Indeed, even if
the $\sim0.5\,$AU region is accretion-friendly, relative velocities
there have values higher than what they should be in
an unperturbed system \citep[that is, approximately the escape velocities of
km-sized bodies, e.g][]{liss93}.
This means that the gravitational focusing factor, which is the cause of
the fast runaway growth mode, is significantly reduced or cancelled.
Growth would thus have to be much slower, maybe orderly or of the
"type II" runaway identified by \citet{kort01}.
In short, planetesimal accretion is not possible in the HZ,
except in the innermost region around 0.5\,AU where
it should be significantly different from what it is around a single star.

However, no conclusions regarding the \emph{presence} of planets in the HZ
can be directly drawn from these results, because of the possible effect
of additional processes not taken into account in our simulations.
In TMS08, we reviewed some of these possible mechanisms, and
concluded that, while some of them might not significantly affect our
conclusions (like bigger "initial" planetesimals or re-accretion in the
subsequent gas-free disc), others might indeed open for a possible presence
of a planet in the terrestrial region.
We reinvestigate here these mechanisms in more details, focusing specifically
on the two most promising scenarios, that of orbital re-phasing during
an extended gas dispersion phase and that of a greater separation for the
early binary.

\subsection{re-phasing during gas dispersion}

\begin{figure}
\includegraphics[width=\columnwidth]{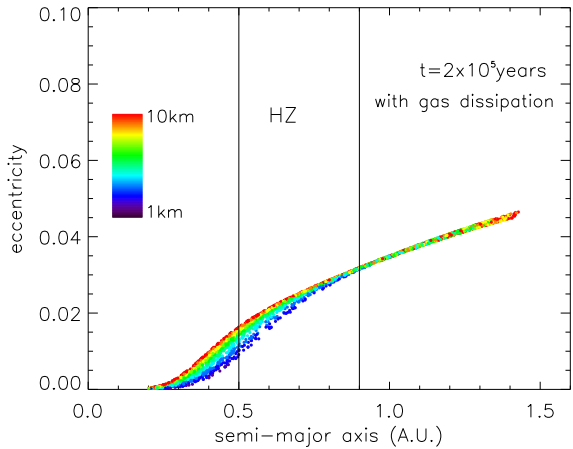}
\caption[]{(e,a) graph, after $t=2\times10^{5}$years, for the run
with gas dissipation. Gas dissipation starts at t=$10^{4}$years
and has a timescale $\tau_{diss}=10^{5}$years (see text for details)
}
\label{eadiss}
\end{figure}

\citet{xie08} have investigated, for the specific case
of the $\gamma$ Cephei binary, what happens to a planetesimal
swarm \emph{during} the gas dispersal period.
They have identified an interesting mechanism:
if the gaseous component is slowly removed from the disc, all
planetesimal orbits are progressively re-phased towards the
same orbits regardless of their sizes. This leads to lower,
possibly accretion-friendly impact velocities.
They concluded that this might allow, under certain circumstances,
for planetesimal accretion to start anew once this realignment is achieved.
We explore this possibility for the present $\alpha\,$Cen\,B case,
assuming the same parameters for gas dispersion as \citet{xie08}: 
$\rho_{g} \propto (t/\tau_{diss})^{-1.5}$,
with $\tau_{diss} = 10^{5}$years. The dissipation is started at $10^{4}$years,
the end of our nominal run, when all orbits have reached their
size-dependent alignment in the HZ. 

Fig.\ref{eadiss} shows the dynamical state of the system after
$2\tau_{diss}=2x10^{5}$years and clearly illustrates the efficient re-phasing
of all orbits, which is almost complete in the $>0.8\,$AU region.
In this outer region, $\langle \Delta v \rangle$
are low enough to allow accretion of all remaining objects. 
However, this encouraging result is undermined by several problems. 
The first one is that, in this accretion-friendly outer region, no
object smaller than $\sim 4\,$km is left after $2\tau_{diss}$, because
the time it takes for most of the gas to be dispersed is long enough
(a few $10^{5}$years) to have a significant inward drift of all smaller
bodies \footnote{This issue had already been identified by
\citet{xie08} but didn't show up in their simulations because of the
numerically imposed re-injection at the outer boundary of all objects
lost at the inner one}.
True, in most of the habitable zone
small planetesimals are still present,
but here full orbital re-phasing is not achieved yet, and the
dynamical environment, even if the situation is slighly improved with
respect to the nominal run, is still globally hostile to accretion.
Only after $\sim 5\tau_{diss}=5\times 10^{5}$years do
$\langle \Delta v \rangle$ become low enough to allow accretion
in the whole HZ, by which time all small objects have here also
been removed.

The second problem is that, even for the larger $\geq 4\,$km
planetesimals, the favourable conditions shown in Fig.\ref{eadiss}
are only reached after an
accretion-hostile transition period of a few $\tau_{diss}$.
The question is then how these objects might survive
this long period during which most impacts will erode them.
These erosion processes cannot be followed with the
N-body models used here, but it is likely that most large
planetesimals will be fragmented into smaller debris which
will be quickly removed from the system
by fast inward drift.

For all these reasons, we do not believe that gas dispersal opens
the possiblity for planetesimal accretion to start anew after a few
$\tau_{diss}$. However, the re-phasing mechanism
during gas dispersal identified by
\citet{xie08} is definitely worth studying in more details.
One crucial, and yet almost unexplored issue is for instance
the relative timing between the planetesimal accretion
phase (in particular its start) and the parallel viscous evolution
of the gas disc. Indeed, as pointed out by
\citet{xie08}, the steady evolution of the gas disc profile,
due to viscous angular momentum redistribution and accretion,
can lead to a gas density decrease of several orders of
magnitudes long before its "definitive" and abrupt dispersion,
at 5 to 10Myrs, resulting from photoevaporation-induced disc
truncature \citep[see e.g., Fig.1 of][]{alex06}.
The question is where (or when) to place the planetesimal accretion
phase in this picture.

\subsection{Wider initial binary separation}

\begin{figure}
\includegraphics[width=\columnwidth]{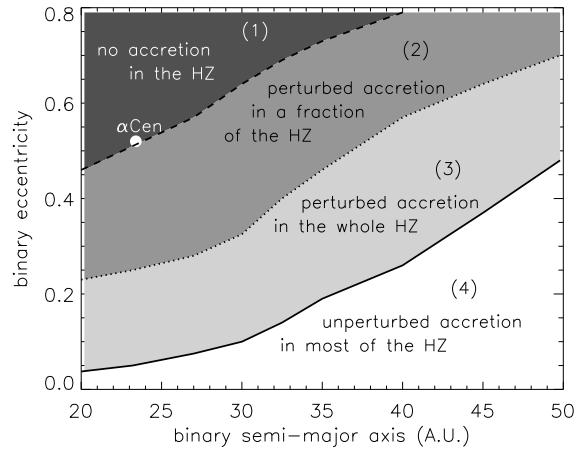}
\caption[]{Dominant collision outcome (erosion, perturbed accretion,
unperturbed accretion, see Fig.\ref{histacc}) in the habitable zone
(0.5-0.9\,AU) around $\alpha$ Centauri B, for different configurations of the
binary's orbit.
}
\label{binpar}
\end{figure}

Most stars are born in clusters. As such they spend
their early life in an environment where
the risk for close stellar encounters is relatively high.
For a binary system, such encounters could
have dramatic effects, breaking it up or significantly
modifying its initial orbit. Conversely, a present
day binary could be what is left of an initially triple (or more)
system \citep{marbar07a,marbar07b}.
As a consequence, there might be an important difference
between the present day orbit of a binary system
and the one it had when planetary formation processes were at play.

In order to have an idea of the statistical chances for an
$\alpha$ Centauri-like binary to have an initially different
orbit, we take as a reference the work of
\citet{malm07}, who studied the dynamical evolution of a typical
open cluster. They found that most binaries of initial
separation $\leq 200\,$AU are not broken up
during the cluster's lifetime. However, a fraction of
these "hard" binaries suffer stellar encounters which alter their orbits.
The net result of these encounters is
to shrink the binaries' orbits, so that separations of young
binaries tend to be larger, with an environment which is thus more
favourable to planet formation. Fig.4 of \cite{malm07}
indicates that $\sim$\,50\% of binaries with separation $\sim\,20$AU
have suffered an orbit-modifying encounter.
There is thus roughly 1 chance in 2 that $\alpha$ Centauri
had a wider separation during the early stages of planet formation
if it formed in a dense cluster.

The crucial question is here which initial orbital configurations
could have allowed planetesimal accretion to occur in the habitable zone
of $\alpha\,$Cen\,B.
We investigate this issue by running a series of 31 test runs for
$20<a_b<50\,$AU and $0<e_b<0.8$ (Fig.\ref{binpar}).
Not surprisingly, we see that present day $\alpha\,$Cen\,B lies
almost at the limit where accretion, although highly perturbed,
becomes possible in the innermost part of the HZ. For the
same $e_b=0.52$ as today, partial perturbed accretion
in the HZ is possible for $23\leq a_b \leq 37\,$AU. In order
to have the HZ become fully accretion-friendly, one
needs $a_b\geq 37\,$AU, while \emph{unperturbed} single star-like
accretion requires $a_b\geq 52\,$AU, more
than twice the present day separation.
The situation gets of course more favourable if one assumes
a lower initial $e_b$ for the binary. As an example, 
with $e_b=0.26$ (half of the present day value), then the whole 
HZ becomes accretion friendly for $a_b\geq 26\,$AU and
allows unperturbed accretion for $a_b\geq 40\,$AU.

These results are of course only a first step
towards understanding this complex issue.
A crucial point is to quantitatively estimate what the probability
for a given $\Delta a_b$ and $\Delta e_b$ "jump" are. 
As an example, how likely is it for a present day $\alpha$ Centauri-like
system to have originated from the accretion-friendly
regions (3) or (4) in Fig.\ref{binpar}? 
The answer to these questions requires very detailed 
and encompassing numerical explorations which have,
to our knowledge, not been performed yet.
\citet{marbar07a,marbar07b} have studied in great detail
the dynamical outcomes of individual encounters, but decoupled
from the general cluster context, while the cluster evolution
study of \citet{malm07} does not give statistical information
about the amplitudes of binary orbital changes.

\section{CONCLUSIONS AND PERSPECTIVES}

We have numerically investigated planetesimal accretion
around $\alpha$ Centauri B. Our main conclusions can be summarized
as follows:
\begin{itemize}
\item Planetesimal accretion is
marginally possible in the innermost parts, $\sim$0.5\,AU,
of the estimated habitable zone. 
Beyond this point, high collision velocities, induced by the coupling
between gas friction and secular perturbations, lead to destructive impacts.
Moreover, even in the $\sim 0.5\,$AU region,
$\langle \Delta v \rangle$ are increased compared to
an unperturbed case. Thus, "classical", single-star like
runaway accretion seems to be ruled out.
\item These results are relatively robust with
respect to the planetesimal size distribution or gas disc profile,
except for a very massive, and probably unrealistic 10xMMSN gas disc.
\item We confirm the conclusions of several previous studies that 
the planetesimal-to-embryo stage is much more affected by binarity
effects than the subsequent embryos-to-final-planets stage.
\item As in \citet{xie08}, we find that
later progressive gas dispersal reduces all $\langle \Delta v \rangle$
to values that might allow accreting impacts. However, we find that the system
has first to undergo a long accretion-hostile transition period during
which most of the smaller planetesimals are removed by inward drift and
most bigger objects are probably fragmented into small debris. Thus,
the positive effect on planetesimal accretion is probably limited.
\item We quantitatively investigate to what extent a wider initial
binary separation, later reduced by stellar encounters in an early
cluster environment, could have favoured planetesimal accretion.
We find that, for a constant $e_b$, an entirely accretion-friendly HZ
requires an initial $a_b\geq 37\,$AU, while normal unperturbed accretion is
only possible for $a_b\geq 52\,$AU. The statistical likehood of such
orbital changes in early open clusters remains to be quantitatively estimated.
\end{itemize}
 
We conclude that, although the presence of planets further out cannot
be fully excluded, the most likely place to look for an habitable terrestrial
planet would be around 0.5\,AU. According to \citet{guedes08}, the 
detection of such a planet could be possible, if its mass is
$\geq 1.8M_{\oplus}$, after 3 years of high cadence observations, provided
that the noise spectrum is white.  
Let us however point out again that the $\sim 0.5\,$AU region around $\alpha\,$Cen\,B
is dynamically perturbed by the binary, so that the formation of a planet
there cannot have followed the standard single-star scenario
(unless the initial binary separation was much larger, see Section 3.2).
The planet formation process in such a perturbed environment,
which could concern many binary systems of intermediate ($\sim\,$20\,AU)
separations, is an important issue which remains to be investigated.
Another crucial issue which remains to be investigated is what happens
\emph{before} the phase studied here. Indeed, our study
implicitly assumes that km-sized planetesimals could form
from smaller grains and pebbles, but the validity
of this assumption has to be critically examined.

{}
\clearpage

\end{document}